\documentclass[]{pasj01}

\Received{}
\Accepted{}
 
\usepackage{url}

\usepackage{lineno}
\newcommand{\icarus}{Icarus}
\begin{document} 
 
\title{ 
The mass determination of TOI-519 b: a close-in giant planet transiting a metal-rich mid-M dwarf}

\author{Taiki \textsc{Kagetani}\altaffilmark{1}}
\email{kagetani@g.ecc.u-tokyo.ac.jp}
\author{Norio \textsc{Narita},\altaffilmark{2,3,4}}
\author{Tadahiro \textsc{Kimura}\altaffilmark{5}}
\author{Teruyuki \textsc{Hirano}\altaffilmark{3,6}}
\author{Masahiro \textsc{Ikoma}\altaffilmark{6,13}}
\author{Hiroyuki Tako \textsc{Ishikawa}\altaffilmark{3,6}}
\author{Steven \textsc{Giacalone}\altaffilmark{7}}
\author{Akihiko \textsc{Fukui}\altaffilmark{2,4}}
\author{Takanori \textsc{Kodama}\altaffilmark{2}}

\author{Rebecca \textsc{Gore}\altaffilmark{7}}
\author{Ashley \textsc{Schroeder}\altaffilmark{7}}
\author{Yasunori \textsc{Hori}\altaffilmark{3,6}}
\author{Kiyoe \textsc{Kawauchi}\altaffilmark{1,4,8}}
\author{Noriharu \textsc{Watanabe}\altaffilmark{1}}
\author{Mayuko \textsc{Mori}\altaffilmark{9}}
\author{Yujie \textsc{Zou}\altaffilmark{1}}
\author{Kai \textsc{Ikuta}\altaffilmark{1}}
\author{Vigneshwaran \textsc{Krishnamurthy}\altaffilmark{3,6}}

\author{Jon \textsc{Zink}\altaffilmark{10}}
\author{Kevin \textsc{Hardegree-Ullman}\altaffilmark{11}}
\author{Hiroki \textsc{Harakawa}\altaffilmark{12}}
\author{Tomoyuki \textsc{Kudo}\altaffilmark{12}}

\author{Takayuki \textsc{Kotani}\altaffilmark{3,6,13}}
\author{Takashi \textsc{Kurokawa}\altaffilmark{6,14}}
\author{Nobuhiko \textsc{Kusakabe}\altaffilmark{3,6}}
\author{Masayuki \textsc{Kuzuhara}\altaffilmark{3,6}}
\author{Jerome \textsc{P. de Leon}\altaffilmark{9}}
\author{John \textsc{H. Livingston}\altaffilmark{3,6,9}}
\author{Jun \textsc{Nishikawa}\altaffilmark{6,13,3}}

\author{Masashi \textsc{Omiya}\altaffilmark{3,6}}
\author{Enric \textsc{Palle}\altaffilmark{4,8}}
\author{Hannu \textsc{Parviainen}\altaffilmark{4,8}}
\author{Takuma \textsc{Serizawa}\altaffilmark{14,6}}
\author{Huan-Yu \textsc{Teng}\altaffilmark{15}}
\author{Akitoshi \textsc{Ueda}\altaffilmark{3,6,13}}
\author{Motohide \textsc{Tamura}\altaffilmark{3,6,9}}

\altaffiltext{1}{Department of Multi-Disciplinary Sciences, Graduate School of Arts and Sciences, The University of Tokyo, 3-8-1 Komaba, Meguro, Tokyo 153-8902, Japan}
\altaffiltext{2}{Komaba Institute for Science, The University of Tokyo, 3-8-1 Komaba, Meguro, Tokyo 153-8902, Japan}
\altaffiltext{3}{Astrobiology Center, 2-21-1 Osawa, Mitaka, Tokyo 181-8588, Japan}
\altaffiltext{4}{Instituto de Astrof\'isica de Canarias, V\'ia L\'actea s/n, E-38205 La Laguna, Tenerife, Spain}
\altaffiltext{5}{Department of Earth and Planetary Science, Graduate School of Science, The University of Tokyo, 7-3-1 Hongo, Bunkyo-ku, Tokyo 113-0033, Japan}
\altaffiltext{6}{National Astronomical Observatory of Japan, 2-21-1 Osawa, Mitaka, Tokyo 181-8588, Japan}
\altaffiltext{7}{Department of Astronomy, University of California Berkeley, Berkeley, CA 94720, USA}
\altaffiltext{8}{Departamento de Astrof\'{i}sica, Universidad de La Laguna (ULL), 38206 La Laguna, Tenerife, Spain}
\altaffiltext{9}{Department of Astronomy, Graduate School of Science, The University of Tokyo, 7-3-1 Hongo, Bunkyo-ku, Tokyo 113-0033, Japan}
\altaffiltext{10}{Department of Astronomy, California Institute of Technology, Pasadena, CA 91125, USA}
\altaffiltext{11}{Steward Observatory, The University of Arizona, Tucson, AZ 85721, USA}
\altaffiltext{12}{Subaru Telescope, 650 N. Aohoku Place, Hilo, HI 96720, USA}
\altaffiltext{13}{Department of Astronomy, School of Science, The Graduate University for Advanced Studies (SOKENDAI), 2-21-1 Osawa, Mitaka, Tokyo, Japan}
\altaffiltext{14}{Tokyo University of Agriculture and Technology, 2-24-16 Nakacho, Koganei, Tokyo 184-8588}
\altaffiltext{15}{Department of Earth and Planetary Sciences, Tokyo Institute of Technology, Meguro-ku, Tokyo, 152-8551, Japan}



\maketitle

\begin{abstract}
We report the mass determination of TOI-519 b, a transiting substellar object around a mid-M dwarf. 
We carried out radial velocity measurements using Subaru / InfraRed Doppler (IRD), revealing that TOI-519 b is a planet with a mass of $0.463^{+0.082}_{-0.088}~M_{\rm Jup}$. We also find that the host star is metal rich ($\rm [Fe/H] = 0.27 \pm 0.09$ dex) and has the lowest effective temperature ($T_{\rm eff}=3322 \pm 49$ K) among all stars hosting known close-in giant planets based on the IRD spectra and mid-resolution infrared spectra obtained with NASA Infrared Telescope Facility / SpeX. 
The core mass of TOI-519 b inferred from a thermal evolution model ranges from $0$ to $\sim30~M_\oplus$, which can be explained by both the core accretion and disk instability models as the formation origins of this planet. 
However, 
TOI-519 is in line with the emerging trend that M dwarfs with close-in giant planets tend to have high metallicity, which may indicate that they formed in the core accretion model. 
The system is also consistent with the potential trend that close-in giant planets around M dwarfs tend to be less massive than those around FGK dwarfs.

\end{abstract}
\clearpage

\section{Introduction}

Thanks to the TESS survey \citep{Ricker+2015}, the number of transiting exoplanets and planetary candidates around M dwarfs has been rapidly increasing. 
However, the number of known short-period ($P<10$ days) giant ($R_p>8~R_\earth$) exoplanets, so-called close-in giant planets, around M dwarfs is still small. In fact, although the number of detected close-in giant planets around FGK dwarfs reaches four hundred, that around M dwarfs is only twelve: Kepler-45 b \citep{Johnson+2012}, HATS-6 b \citep{Hartman+2015}, NGTS-1 b \citep{Bayliss+2018}, HATS-71 b \citep{Bakos+2020}, TOI-530 b \citep{Gan+2022}, HATS-74A b, HATS-75 b \citep{Jord+2022}, TOI-3629 b, TOI-3714 b \citep{Canas+2022}, TOI-3757 b \citep{Kanodia+2022}, TOI-5205 b \citep{Kanodia+2023}, and TOI-3235 b \citep{Hobson+2023}. The core accretion theory of planet formation predicts that it is hard to form Jupiter-sized giant planets around low-mass stars (\cite{Laughlin+2004}; \cite{Ida+2005}; \cite{Kennedy+2008}), which is consistent with observations to date. 
The amount of material typically found in protoplanetary disks around low-mass stars is not so massive (e.g., \cite{Ansdell+2017}), which supports the prediction of the core accretion theory.
However, it is still unclear if giant planets around low-mass stars form in this way, or if a competing theory, which is the disk instability theory \citep{Cameron+1978, Boss1997}, offer a better explanation.
Increasing the sample size and investigating the correlations between the planetary and stellar properties (e.g., planetary mass v.s. stellar metallicity)  are required to reveal the formation mechanisms of close-in giant planets around low-mass stars.

TOI-519 b is a Jupiter-sized companion transiting an M dwarf, which was validated with multi-color transit photometry with the MuSCAT2 instrument \citep{Narita+2019} and phase curve analysis of the TESS data in \citet{Parviainen+2021}. However, \cite{Parviainen+2021} did not confirm if TOI-519 b is a giant planet or a brown dwarf because they were only able to place an upper limit on the companion mass of $M_p<14M_{\rm Jup}$ at $2 \sigma$.
We carried out the radial velocity (RV) measurements with the Subaru / InfraRed Doppler (IRD) in order to discriminate whether TOI-519 b is a giant planet or a brown dwarf by measuring its mass. We also refine the stellar parameters such as the radius, mass, luminosity, effective temperature, and metallicity using both IRD and NASA Infrared Telescope Facility (IRTF) / SpeX.

This paper is structured as follows: section 2 presents the spectroscopic observations of IRD and SpeX. The analysis and results are presented in section 3. Section 4 provides the discussion of the possible planet formation scenario and interior of TOI-519 b, and the characteristics of TOI-519 b compared with other close-in giant planets. We end with a conclusion in section 5.



\section{Observations}

\subsection{Subaru IRD}\label{subsec:IRD}

The RV data of TOI-519 b were obtained using IRD instrument installed on the 8.2m Subaru telescope (\cite{Tamura+2012}; \cite{Kotani+2018}), which covers the near-infrared wavelength range from 930 nm to 1740 nm with a resolution of $\approx$70000.  We carried out these observations for thirteen nights 
between UT 2019 November 9 to 2022 January 24 under the Subaru-IRD TESS intensive follow-up program (ID: S19A-069I, S20B-088I, S21B-118I). We obtained in total 25 data points, of which nine were obtained in both $YJ-$band and $H-$band, while the others were observed in only $H-$band due to a technical problem on the $YJ-$band detector. In order to get high S/N, we set the  integration time to $1800-2100$ sec for each exposure depending on the observing condition. 
The extracted one-dimensional spectra have 
the S/N ratios of $15-28$ and $15-53$ per pixel at 1000 nm and 1600 nm, respectively.
The median S/N ratios of 22 (1000 nm) and 47 (1600 nm) for those frames were consistent with the expected S/N values for very faint stars; for a $J=12.8$ mag star, IRD would be able to achieve an S/N ratio of $\approx 25$ at 1000 nm with a $30$-minute integration. 
We also observed at least one telluric standard star (A0 or A1 star) on each night to correct for the telluric lines in extracting the template spectrum for the RV analysis. 

\subsection{IRTF SpeX}

We observed TOI-519 on UT 2019 April 19 with the SpeX spectrograph \citep{2003PASP..115..362R, 2004SPIE.5492.1498R} on the 3m NASA IRTF. Our data was collected in the SXD mode using the $0\farcs3 \times 15\arcsec$ slit and covered a wavelength range of $0.7-2.55$ $\mu$m.


\section{Analysis and Results}


\subsection{Analysis of IRD spectra}

\subsubsection{IRD Data Reduction and RV Measurements}

Using the wavelength-calibrated, one-dimensional (1D) spectra for 
TOI-519 as well as the simultaneous reference spectra of the laser-frequency comb, 
we extracted precise RVs for TOI-519. 
The RV pipeline for IRD is described in detail in \citet{Hirano+2020};  in short, we first derived 
a template spectrum of the target, which is free from the instrumental broadening and 
telluric features, and relative RVs for individual frames are measured with respect to this 
template spectrum by the forward modeling technique. 
Since we only obtained $H$-band spectra in 2022 January, 
we generated the stellar templates for the $YJ-$band and $H-$band separately, using
different numbers of frames;
we combined 9 frames taken between 2019--2021 for the $YJ-$band while we used 16 frames secured in 2019-2022, which were selected based on the S/N ratios. Some of the spectra in this analysis may have suffered from the detector's persistence (see Section \ref{sec:orbitalfit}), but its impact had to be negligible given that the stellar template was derived by combining only high S/N frames by median. 
Individual RVs were also derived separately for  the $YJ-$ and $H-$bands.
We also measured the systemic RV for TOI-519 by fitting relatively deep atomic lines by Gaussian and comparing their centers to the theoretical positions. 
We obtained $\mathrm{RV}=26.89\pm 0.57$ km s$^-{1}$, whose uncertainty was determined
based on the scatter of the line positions.

Activity indicators including the line full width at half maximum (FWHM) and line asymmetry indices have been studied for IRD spectra \citep{2022PASJ...74..904H}, but in their derivations only spectral regions free from telluric lines in the $Y$-band have been used. For TOI-519, we did not derive the activity indicators as most of the frames in Table \ref{tab:RVdata} were obtained with the $H$-band detector alone.

\begin{table}
  \tbl{stellar parameters of the TOI-519.}{%
  \begin{tabular}{cccc}
      \hline
      Parameter & Value & Reference \\ 
      \hline
      $R_{\star}$ ($R_{\odot}$)   & $0.373^{+0.020}_{-0.088}$ & (1) \\
      $R_{\star}$ ($R_{\odot}$)   & $0.350 \pm 0.010$ & (2) \\
      $M_{\star}$ ($M_{\odot}$)   & $0.369^{+0.026}_{-0.097}$ & (1) \\
      $M_{\star}$ ($M_{\odot}$)   & $0.335 \pm 0.008$ & (2) \\ 
      $\log g_{s}$ (cgs)   & $4.87 \pm 0.03$ & (2) \\
      $T_{\rm eff}$ (K)   & $3350^{+100}_{-200}$ & (1) \\
      $T_{\rm eff}$ (K)   & \ $3322 \pm 49$ & (2) \\
      $L_{\star}$ ($L_{\odot}$)   & $0.0141 \pm 0.0006$ & (2) \\
      $\rm [Fe/H]$ (dex)   & $0.27 \pm 0.09$ & (2) \\
      Parallax (mas) & $8.6807 \pm 0.0373$ & (3) \\
      Distance (pc) & $115.2 \pm 0.5$ & (2) \\
      RV ($\rm km~s^{-1}$)   & $26.89 \pm 0.57$ & (2) \\
      $U$ ($\rm km~s^{-1}$)   & $-36.35 \pm 0.29$ & (2) \\
      $V$ ($\rm km~s^{-1}$)   & $-11.89 \pm 0.49$ & (2) \\
      $W$ ($\rm km~s^{-1}$)   & $-6.14 \pm 0.10$ & (2) \\
      Age (Gyr)    & $0.8 - 10$ & (2) \\
      Age (Gyr)    & $6.00 \pm 2.59$ & (4) \\
      \hline
      \end{tabular}}\label{tab:stellar parameter}
\begin{tabnote}
(1) \citet{Parviainen+2021}. 
(2) This work.  
(3) Gaia DR3.
(4) \citet{Gaidos+2023}.
\end{tabnote}
\end{table}

\subsubsection{Estimation of Stellar Parameters}
To derive the effective temperature $T_{\mathrm{eff}}$ and the abundances of individual elements [X/H], we used a combined IRD spectrum. Here the telluric features were removed as was done for the template spectrum for the RV measurements but the instrumental broadening was not deconvolved. It is because the deconvolution of the instrumental profile amplifies the random noise due to the low S/N ratio of the original data, especially in the YJ-band which we used for the estimation of stellar parameters.

The analyses are based on the equivalent width (EW) comparison of individual absorption lines between synthetic spectra and the observed one. 
The synthetic spectra were calculated with a 1-D LTE radiative transfer code, based on the same set of assumptions as in the model atmosphere program of \citet{1978A&A....62...29T} such as plane-parallel geometry, radiative and hydrostatic equilibrium, and chemical equilibrium.

For $T_{\mathrm{eff}}$ estimation, 29 FeH molecular lines at $990-1012$ nm were used following the procedure described in \citet{2022AJ....163...72I}.
For abundances, 28 atomic lines in total created by neutral atoms of Na, Mg, Ca, Ti, Cr, and Fe and singly ionized Sr were employed.
The lines were selected that were not blended with other lines and whose continuum levels could be determined at the surrounding wavelengths. For elements with a sufficient number of absorption lines, we removed lines that were too deep or too broad, which could cause low sensitivity to the elemental abundance.
To determine the EW of each line, we fitted a Gaussian profile to the FeH line, while a synthetic spectrum to the atomic line.
More details of the abundance analysis are outlined in \citet{2020PASJ...72..102I}, while not all the lines they used were available in this work due to the low S/N ratio.

We derived a provisional $T_{\mathrm{eff}}$ adopting the solar abundances, and then, we determined the individual abundances of the seven elements [X/H] adopting the provisional $T_{\mathrm{eff}}$. 
Subsequently, we redetermined $T_{\mathrm{eff}}$ adopting the elemental abundances resulting from the first analysis as the input.
These determinations of $T_{\mathrm{eff}}$ and abundances were iterated alternately until the final results converged within the error margins. 
We obtained [Fe/H] = $0.46 \pm 0.18$ dex and $T_{\mathrm{eff}} = 3229 \pm 100$ K. 
We also calculated an error-weighted average of the final abundances of the seven individual elements to obtain [M/H] = $0.48 \pm 0.08$ dex. 

Note that the $T_{\mathrm{eff}}$ error has systematic errors ($\sigma_{\rm sys}\sim$100 K) as well as statistical error, which was discussed in  \citet{2022AJ....163...72I}. The statistical error ($\sigma_{\rm stat}=22$ K) is the standard deviation of the estimates from individual lines divided by the square root of the number of the lines ($\sigma / \sqrt{N}$). We adopted 100 K for the $T_{\mathrm{eff}}$ error because systematic error is sufficiently dominant compared to statistical error.


We estimated the barycentric Galactic space velocities $(U, V, W)$ as listed in table \ref{tab:stellar parameter} using the astrometric information in Gaia DR3 \citep{Gaia+2022}, the distance derived from the inverse of the parallax, and the systemic RV obtained from the IRD spectra.  According to \citet{Bensby+2014}, the probabilities that the system belongs to the Galactic thin disk, thick disk, and halo are calculated to be $P(\rm thin) = 98~\%$, $P(\rm thick) = 2~\%$, and $P(\rm halo) < 0.1~\%$, respectively. This result and figure 2 in \citet{Xiang+2022}, which showed stellar age-metallicity relation, indicate that the age of TOI-519 is less than 10 Gyr. In addition, based on the value of pseudo-equivalent width of H$\alpha$ emission \citep{Parviainen+2021} and relation between H$\alpha$ emission and stellar age \citep{Kiman+2021}, we placed the lower limit of 0.8 Gyr on the stellar age. 
Note that \citet{Gaidos+2023} recently estimated the age of this system to be $6.00 \pm 2.59$ Gyr by gyrochronology.

\begin{figure*}
    \centering
    \includegraphics[width=\linewidth]{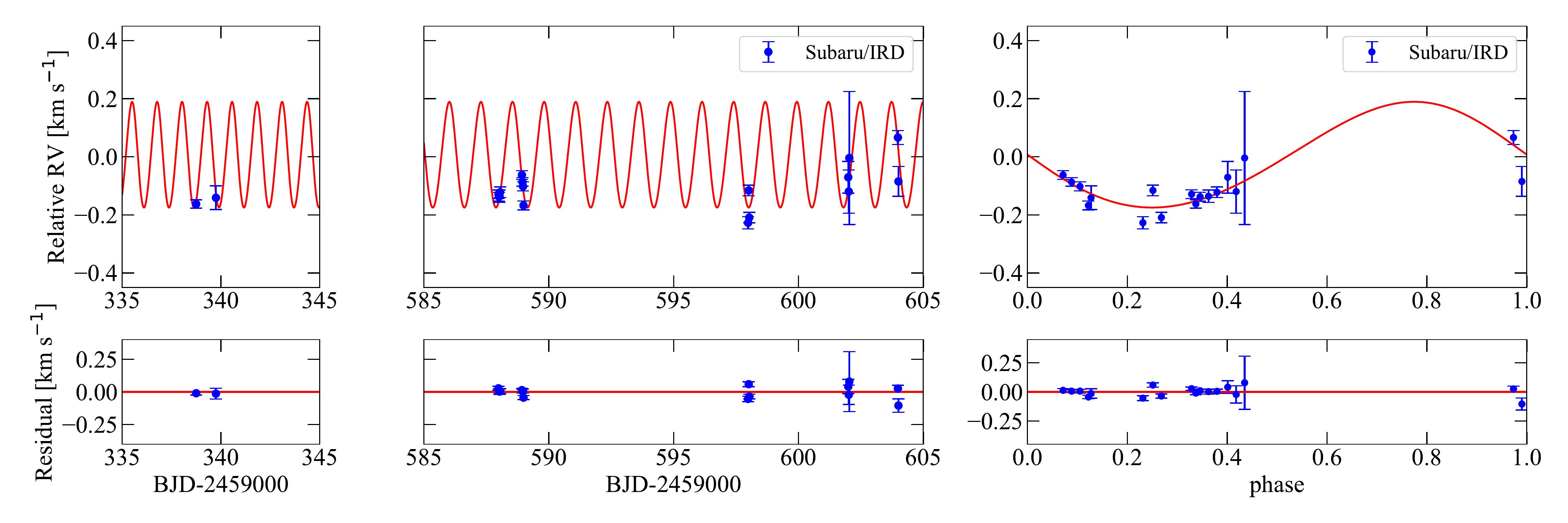}
    \caption{Relative RVs of TOI-519 are presented in the upper panel and bottom panels show residuals. Blue plots show Subaru/IRD data, and red lines indicate the best fitting model. Left and center panel shows the time series and right panel shows the phase-folded. }
    \label{fig:RVfit}
\end{figure*}

\subsection{Analysis of SpeX spectra}

The data was reduced using the \texttt{Spextool} pipeline \citep{2004PASP..116..362C}. After reducing, RV-correction was applied to our spectrum using \texttt{tellrv} \citep{2014AJ....147...20N}. We then calculated the metallicity of TOI-519 following the relations defined in \citet{2013AJ....145...52M} for cool dwarfs with spectral types between K7 and M5. In performing this calculation, we opted to only use the $Ks$-band spectrum, which \citet{2019AJ....158...87D} found to be more reliable than those extracted by the $H$-band spectrum and less affected by telluric contamination. Our analysis yielded metallicities of [Fe/H] = $0.21 \pm 0.10$ dex and [M/H] = $0.11 \pm 0.10$ dex.

To determine the effective temperature of TOI-519, we followed the procedure outlined in section 4.3 of \citet{2019AJ....158...87D}. We began by calculating the absolute $K_s$ and $J$ magnitudes ($M_{K_s}$ and $M_J$) using 2MASS \citep{2MASS2006} and the stellar parallax presented in Gaia DR3 \citep{Gaia+2022}. We then calculated the radius and luminosity of the star using the relations outlined in \citet{Mann+2015}. Specifically, radius was determined using the $R_\star - M_{K_s} - {\rm [Fe/H]}$ relation in table 1 of \citet{Mann+2015} and luminosity was determined using $M_J$ and the metallicity-dependent $r-J$ bolometric correction in table 3 of \citet{Mann+2015}, where the $r$ magnitude of the star ($r = 16.038 \pm 0.205$) was found in the Carlsburg Meridian Catalogue \citep{2014AN....335..367M}. Lastly, we calculated effective temperature using the Stefan-Boltzmann relation, finding $T_{\rm eff} = 3353 \pm 59 $ K. To ensure that the stellar parameters derived from these methods are robust, we repeated the luminosity calculation using the metallicity-dependent $V-J$ bolometric correction in table 3 of \citet{Mann+2015}, adopting $V$ from \citet{Parviainen+2021}. Our derived luminosity and $T_{\rm eff}$ values using the two different bolometric corrections agreed with one another to within $1\sigma$. Hereafter, we adopt the values obtain with the $r-J$ relation.

We adopted the weighted mean of the two respective measurements of IRD ([Fe/H] = $0.46 \pm 0.18$ dex) and SpeX ([Fe/H] = $0.21 \pm 0.10$ dex) for [Fe/H] and three respective measurements of IRD ($T_{\rm eff} = 3229 \pm 100 $ K), \citet{Parviainen+2021} ($T_{\rm eff} = 3350^{+100}_{-200} $ K), and Mann relationship ($T_{\rm eff} = 3353 \pm 59 $ K) for $T_{\rm eff}$, specifically, [Fe/H] = $0.27 \pm 0.09$ dex and $T_{\rm eff} = 3322 \pm 49$ K. We recalculated the $R_\star$ through empirical relation using equation (5) in \citet{Mann+2015} and $M_\star$ using equation (5) in \citet{Mann+2019}. As a result, the stellar radius and mass are derived to be $R_\star = 0.350 \pm 0.010~R_\odot$ and $M_\star = 0.335 \pm 0.008~M_\odot$, respectively. We also update the planetary radius to $R_p = 1.03 \pm 0.03~R_{\rm Jup}$ using the value of $R_p/R_\star$ in \citet{Parviainen+2021} and other physical parameters of the planet (see table \ref{tab:derived parameter}).

\begin{table}
  \tbl{RVs of TOI-519}{%
  \begin{tabular}{cccc}
      \hline
      \hline
      BJD & RV ($\rm km~s^{-1}$) & $\sigma$~($\rm km~s^{-1}$) & S/N\footnotemark[$a$]  \\ 
      \hline
      2459338.74458644 & -0.163 & 0.014 & 52.6  \\
      2459339.74415244 & -0.141 & 0.041 & 26.1  \\
      2459587.98455199 & -0.129 & 0.015 & 51.7  \\
      2459588.00607961 & -0.138 & 0.015 & 49.2  \\
      2459588.02744583 & -0.136 & 0.021 & 39.9  \\
      2459588.04902426 & -0.122 & 0.018 & 42.4  \\
      2459588.92422570 & -0.063 & 0.015 & 50.6  \\
      2459588.94566140 & -0.087 & 0.015 & 50.3  \\
      2459588.96706379 & -0.102 & 0.015 & 51.7  \\
      2459588.98840808 & -0.167 & 0.015 & 49.0  \\
      2459597.98299815 & -0.227 & 0.021 & 45.6  \\ 
      2459598.00844307 & -0.116 & 0.018 & 45.8  \\
      2459598.02980661 & -0.210 & 0.018 & 46.6  \\
      2459601.99374490 & -0.071 & 0.055 & 22.9  \\
      2459602.01513202 & -0.120 & 0.074 & 19.8  \\
      2459602.03649085 & -0.004 & 0.229 & 14.6  \\
      2459603.98294345 & 0.066  & 0.024 & 36.2  \\
      2459604.00433421 & -0.085 & 0.051 & 23.6  \\
      \hline
      \end{tabular}}\label{tab:RVdata}
\begin{tabnote}
\footnotemark[$a$] S/N are values at 1600 nm  \\ 
All data exposure times are 1800 s \\
\end{tabnote}
\end{table}

\subsection{Orbital Fits to the RV Data}\label{sec:orbitalfit}

It has been revealed in IRD data that, when a faint object like TOI-519 is observed just after a bright object is exposed, the detector persistence has a significant impact on data (\cite{Hirano+2020a}; \cite{Hirano+2020b}). We checked the raw IRD data for both TOI-519 and the previously exposed object, and excluded the data with a significant effect of detector persistence from the RV fit. 
In addition, since most $H-$band data have considerably better S/N than the $YJ-$band data, we used only $H-$band data. The eighteen RVs which we use are listed in table~\ref{tab:RVdata}. All adopted data has an exposure time of 1800 s.

The relative RVs of TOI-519 measured with Subaru IRD are shown in figure \ref{fig:RVfit}. 
In order to constrain the companion's mass, we carried out an RV fit with five parameters: the RV semi-amplitude $K$, the RV zero point $V_0$, two eccentricity components, $\sqrt{e}~\mathrm{sin}~\omega$, $\sqrt{e}~\mathrm{cos}~\omega$, and the RV jitter $\sigma_{\rm jit}$. We fixed other orbital parameters: the orbital period $P$, the transit center $t_c$, and the inclination $i$ at the values in \citet{Parviainen+2021}.  
First, we determined the maximum likelihood value of the model using \texttt{scipy.optimize}, and then we performed an MCMC analysis using \texttt{emcee} \citep{Forman-Mackey+2013}. We adopted uniform priors on all the parameters (see table~\ref{tab:orbital parameter}). 
Our analysis of the RVs yielded a semi-amplitude of $K=0.182^{+0.032}_{-0.034}~\rm km~s^{-1}$, which corresponds to the companion's mass of $0.463^{+0.082}_{-0.088}~M_{\rm Jup}$. We determined the mass of the companion at better than $5\sigma$ level and confirm that TOI-519 b is a giant planet rather than a brown dwarf. We placed eccentricity $3~\sigma$ upper limit of $e<0.33$. We also estimated the circularization timescale using equation (4) in \citet{Jackson+2008}. When we adopt the value of tidal quality factor for $Q_p=10^5-10^8$, circularization timescale is $<0.8$ Gyr.
It seems unlikely that there is any remaining eccentricity, but even if there is, it would be small.





\begin{table}
  \tbl{orbital parameters.}{%
  \begin{tabular}{cccc}
      \hline
      Parameter & prior & Value \\ 
      \hline
      $P$ (days) & Fixed & $1.2652328 \pm 5\times10^{-7}$  \\
      $K$ ($\rm km~s^{-1}$) & $\mathcal{U}$(0,$\infty$) & $0.182^{+0.032}_{-0.034}$\\
      $\sigma_{\rm jit}$ ($\rm km~s^{-1}$) & $\mathcal{U}$(0,5) & $0.029^{+0.011}_{-0.008}$\\
      $\sqrt{e}~\mathrm{cos}~\omega$ & $\mathcal{U}$(-1,1) & $0.0^{+0.2}_{-0.1}$ \\
      $\sqrt{e}~\mathrm{sin}~\omega$ & $\mathcal{U}$(-1,1) & $0.1 \pm 0.2$  \\
      \hline
      \end{tabular}}\label{tab:orbital parameter}
\begin{tabnote}
\end{tabnote}
\end{table}

\begin{table}
  \tbl{derived parameters of the TOI-519 b.}{%
  \begin{tabular}{cccc}
      \hline
      Parameter & Value & Reference \\ 
      \hline
      $R_{p}$ ($R_{\rm Jup}$)   & $1.06 \pm 0.17$ & (1) \\
      $R_{p}$ ($R_{\rm Jup}$)   & $1.03 \pm 0.03$ & (2) \\
      $i$ (deg)   & $88.9 \pm 0.4$ & (3) \\
      $e$  & $0.06^{+0.09}_{-0.04}$, $3~\sigma<0.33$ & (2) \\
      $M_{p}$ ($M_{\rm Jup}$)   & $0.463^{+0.082}_{-0.088}$ & (2) \\
      $\rho_{p}$ ($\rm g~cm^{-3}$)   & $0.56 \pm 0.11$ & (2) \\
      $\log g_{p}$ (cgs)  & $3.05 \pm 0.09$ & (2) \\
      $a$ (au)  & $0.0159 \pm 0.0001$ & (2) \\
      $T_{\rm eq}$ (K)  & $687 \pm 14$ & (2) \\
      \hline
      \end{tabular}}\label{tab:derived parameter}
\begin{tabnote}
(1) Rederive using $k_{\rm true}$ and $R_{\star}$ as reported by \citet{Parviainen+2021}.
(2) This work.  
(3)\citet{Parviainen+2021}. \\
\end{tabnote}
\end{table}

\section{Discussion}

\subsection{Origin and interior of TOI-519~b}

The leading hypothesis for the formation of gas giant planets is that runaway gas accretion takes place once a solid core becomes massive enough (i.e., the core accretion hypothesis; \cite{Mizuno1980}; \cite{Bodenheimer+1986}). 
As far as hot Jupiters orbiting Sun-like stars are concerned, the core accretion hypothesis is supported by the well-known positive correlation between planet occurrence and stellar metallicity (e.g., \cite{Fischer+2005, Johnson+2010, Maldonado+2012, Schlaufman+2014, Adibekyan+2019,Osborn+2020}). 
Such a correlation is consistent with the idea that there are enough solids available to form massive cores in protoplanetary disks around metal-rich stars. 
In the same light, around low-mass stars, giant planet formation is less likely because of insufficient amounts of materials (e.g., \cite{Ida+2005}) although there are results from RV studies that have investigated the correlation between giant planet occurrence and steller metallicity (e.g., \cite{Neves+2013,Maldonado+2020}). 
\citet{Burn+2021} also shows that giant planets can not form around low-mass stars ($M_s<0.5~M_\odot$), which reflect the limitations in our understanding of disk properties, migration rates, and core accretion for M dwarfs; instead, the direct formation through disk instability may bring about gas giant planets around low-mass stars (\cite{Cameron+1978}; \cite{Boss1997}). 
Given the high value of [Fe/H] of the host star (see table~\ref{tab:stellar parameter}), either origin could be considered for TOI-519~b. 
A crucial constraint is the mass of the core or metals contained in the planetary interior.

Our analysis above reveals that TOI-519~b is similar in radius to but smaller in mass than Jupiter. While Saturn is inferred to contain metals of $19.1\pm1.0~M_\oplus$ 
in its interior \citep{Mankovich+2021}, such a low mean density of TOI-519~b (= $0.56\pm0.11$~g\,cm$^{-3}$) relative to Saturn's (= 0.69~g\,cm$^{-3}$) implies that this planet contains less metals. 
For quantitative estimate of the metal content, however, one needs to consider the effects of age and stellar irradiation and the efficiency of planetary cooling.

We simulate the thermal evolution of TOI-519~b by standard modeling for the structure of a gas giant planet with an irradiated radiative H/He atmosphere on top of a fully convective H/He envelope on top of a central ice/rock core (e.g., \cite{Kurosaki+2017}). 
For the atmospheric structure, we use the analytical solution from \citet{Matsui+1986} with the equilibrium temperature of 700~K, which is almost the same as that of TOI-519 b, and the albedo of 0.1; the atmosphere/envelope boundary is assumed to locate at a pressure of 1~kbar. 
For the envelope structure, we use \citet{Saumon+1995}'s equation of state for the H-He mixture. 
We do not solve the structure of the core, but use the mass-radius relationship theoretically derived by \citet{Fortney+2007}; the ice/rock ratio is assumed to be unity.

\begin{figure}
    \centering
    \includegraphics[width=\columnwidth]{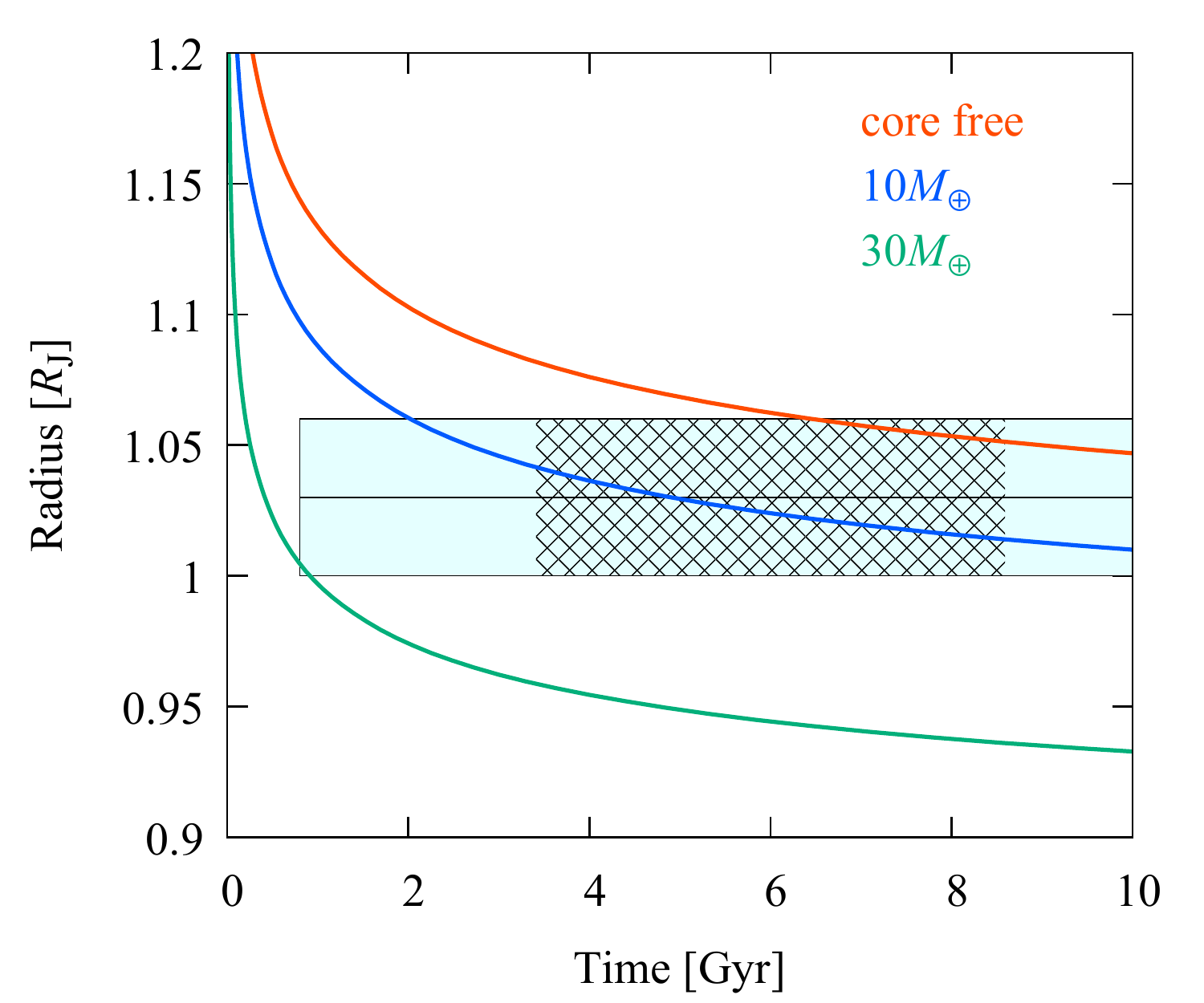}
    \caption{Simulated thermal evolution of TOI-519~b. 
    The planetary radius in $R_{\rm Jup}$ is plotted as a function of age in Gyr. 
    The planet is assumed to consist of a central ice/rock core plus a H/He envelope; the region below 1~kbar is considered as the atmosphere separately from the isentropic envelope. 
    In the calculations shown with solid lines, the elemental abundances in the atmosphere are assumed solar.
    The lines are color-coded according to the assumed core mass (see legend). 
    The light-blue area indicates the measured radius of TOI-519~b and the age of TOI-519 derived in this study. The hatched area indicates the age of TOI-519 derived by \citet{Gaidos+2023} (see table~\ref{tab:stellar parameter}).
    }
    \label{fig:Rp_time}
\end{figure}

The simulated thermal evolution is shown in figure~\ref{fig:Rp_time}, where the planet radius is plotted as a function of age for different choices of the core mass, $M_{\rm core}$, which is compared with the measured radius (light-blue area). 
In the calculations indicated by solid lines, we have assumed the atmosphere with the solar composition.
The presence of a core of $M_{\rm core} \lesssim$~30~$M_\oplus$ accounts for the measured radius. 
A core mass is $M_{\rm core} \lesssim$~20~$M_\oplus$ when the age derived by \citet{Gaidos+2023} is taken into account.
Standard core accretion models (e.g., \cite{Pollack+1996}; \cite{Ikoma+2000}) show that $\sim$5--20~$M_\oplus$ of a core suffice to trigger runaway gas accretion. 
Thus, the inferred core mass does not conflict with the core-accretion origin of this planet. 
Since the dust mass of protoplanetary disks around M dwarfs are not so massive ($M_{dust} \lesssim~25~M_\oplus$) (e.g., \cite{Ansdell+2017}), the formation of a core as heavy as $20 - 30~M_\oplus$ around M dwarfs would be very difficult, though not impossible because the total disk solid mass could be much higher when considering the amount of planetesimals. 
Since TOI-519 b is part of a small population of close-in giant planets recently increasing around mid-M dwarfs, this planet could be a rare planet with an unusual core mass of $\sim20-30~M_\oplus$.
Meanwhile, within the error range of planet radius, one cannot rule out the possibility of a metal-free planet and, thus, the disk-instability origin, although rather massive disks (with tens of percent of host-stellar mass) are needed for disk instability around low-mass stars \citep{Mercer+2020}. 
Comparing the radius and mass relation models predicted for each equilibrium temperature for core-free giant planets around FGK dwarfs (see figure 2 in \cite{Throngren+2018}), the radius and mass relation of TOI-519 b is mostly consistent with them and almost all other close-in giant planets around M dwarfs as well, which indicate that the population would be appear to be consistent with expectations from hot and warm Jupiters orbiting FGK dwarfs. 
This is also suggest that the core mass of TOI-519 b is relatively small.

More precise determination of the stellar age and planet's radius, and measurement of atmospheric composition will help us better understand the interior and origin of TOI-519~b.

\begin{figure}
    \centering
    \includegraphics[width=\columnwidth]{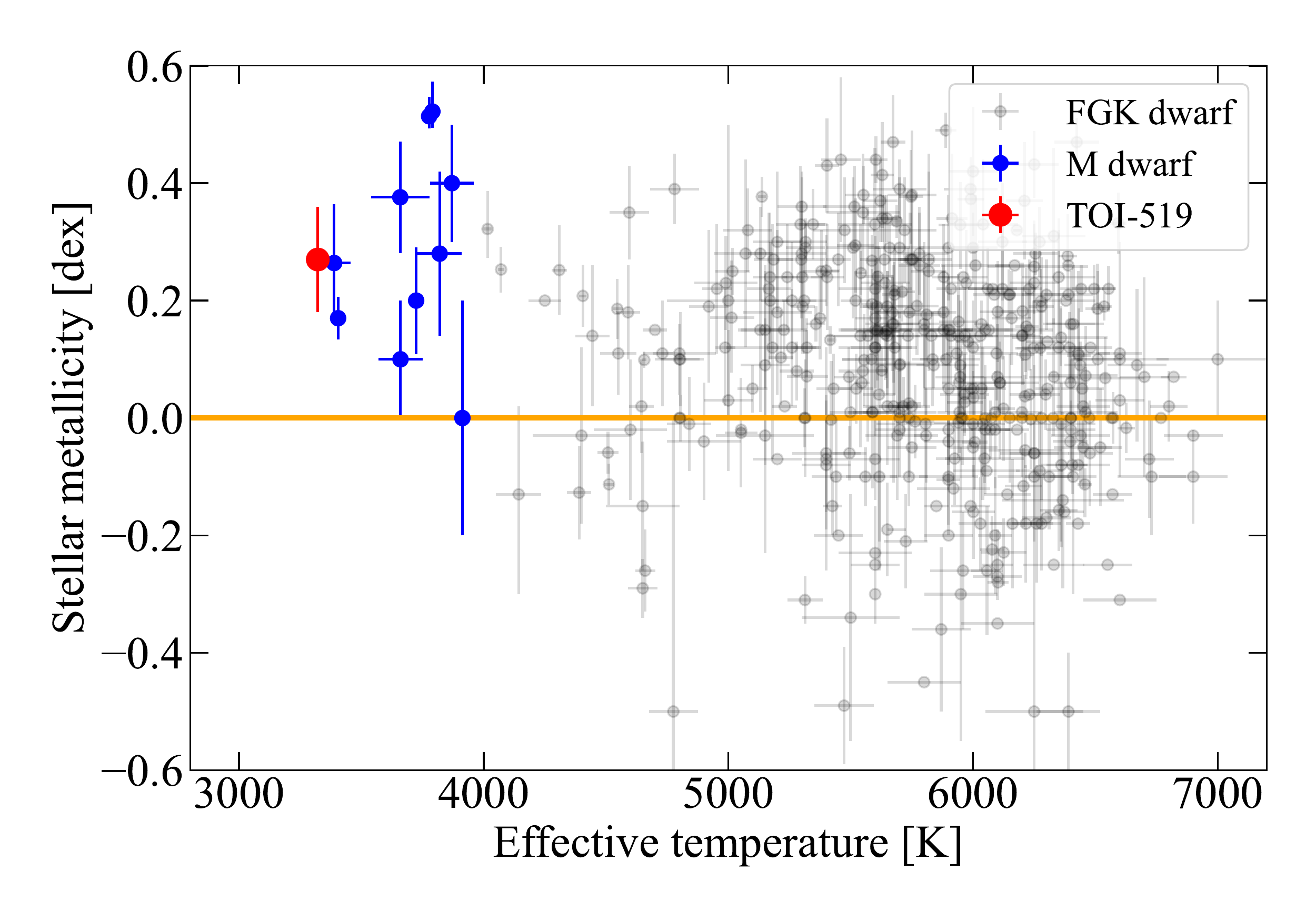}
    \caption{The distribution of the effective temperature and metallicity of FGKM dwarfs hosting close-in giant planets discovered so far. The gray, blue, and red plots represent FGK dwarfs, M dwarfs, and TOI-519, respectively. The orange line shows the solar metallicity ([Fe/H] = 0). NGTS-1 b \citep{Bayliss+2018} and TOI-5205 b \citep{Kanodia+2023} are not plotted because metallicity were not measured precisely.}
    \label{fig:Teff-metal}
\end{figure}

\begin{figure}
    \centering
    \includegraphics[width=\columnwidth]{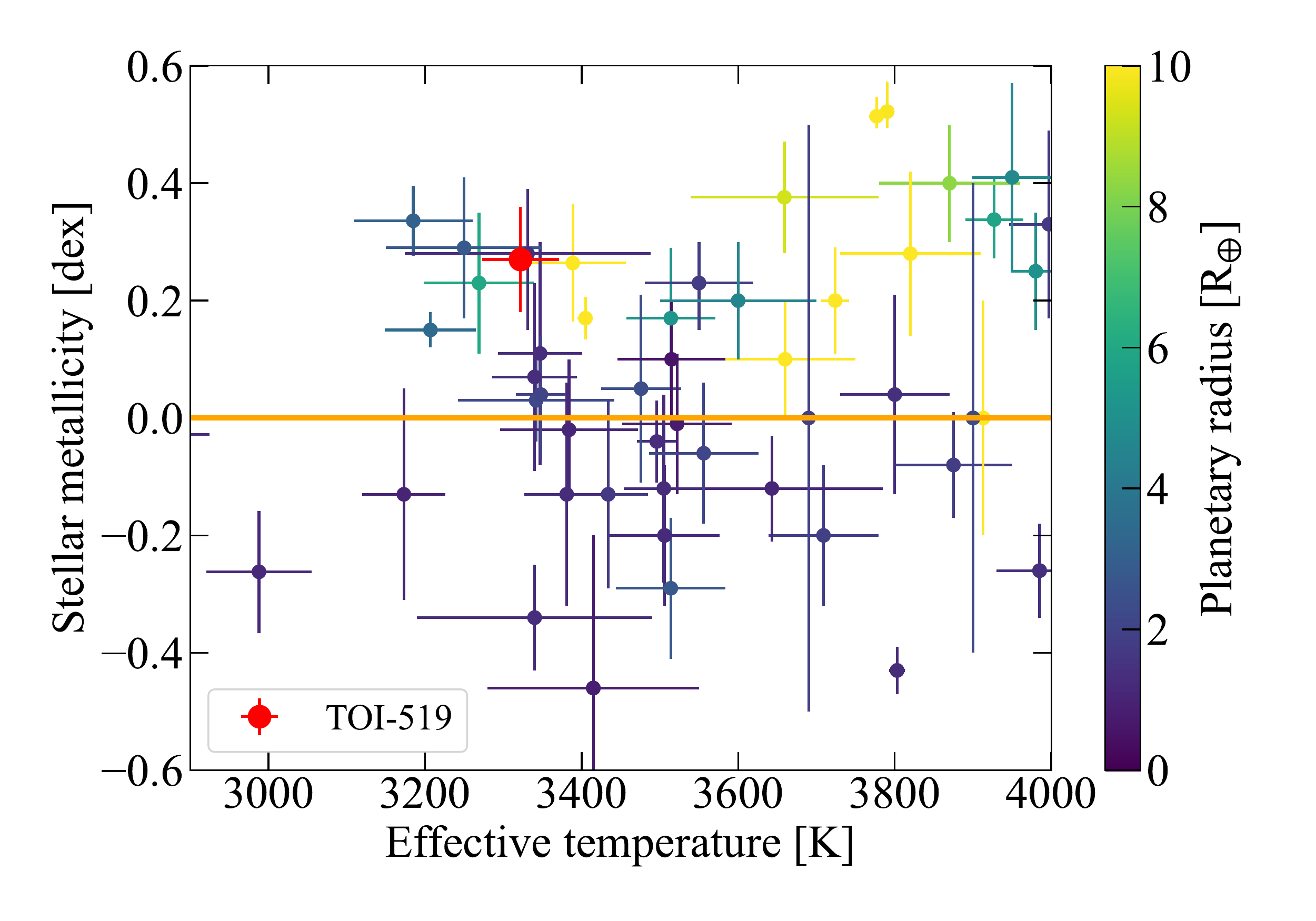}
    \caption{The distribution of the effective temperature and metallicity of M dwarfs hosting close-in planets discovered so far. Colors represent the planetary radius. The red plot shows TOI-519. The orange line shows the solar metallicity ([Fe/H] = 0). NGTS-1 b \citep{Bayliss+2018} and TOI-5205 b \citep{Kanodia+2023} are not plotted because metallicity were not measured precisely.}
    \label{fig:Teff-metal_Mdwarfs}
\end{figure}

\subsection{Comparison with other close-in giant planets}

TOI-519 b is the thirteenth close-in giant planet ($P<10$ days and $R_p>8~R_\earth$) that orbits an M dwarf. Although the origin of TOI-519 b cannot be concluded with the current observations (see section 4.1), one might see correlations between the planetary and stellar properties in the M dwarf systems with giant planets, if the planets formed in the same mechanism. In the framework of the core accretion theory, stellar metallicity is an important factor as described in section 4.1.

\begin{figure}
    \centering
    \includegraphics[scale=0.35]{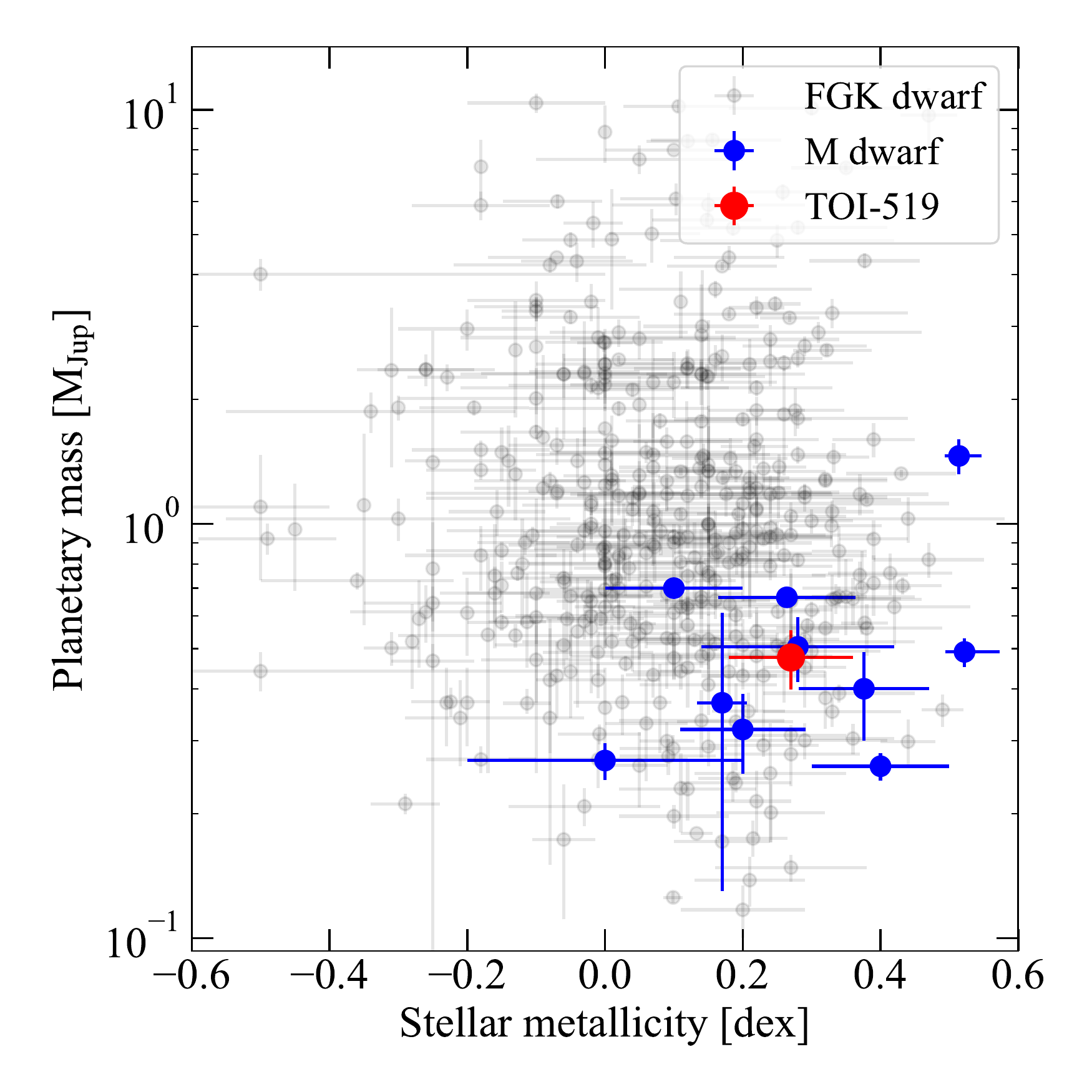}
    \caption{The distribution of stellar metallicity and planetary mass. Plotted system and their plot colors are the same as figure~\ref{fig:Teff-metal}. NGTS-1 b \citep{Bayliss+2018} and TOI-5205 b \citep{Kanodia+2023} are not plotted because metallicity were not measured precisely.}
    \label{fig:metal-Mp}
\end{figure}

Figure~\ref{fig:Teff-metal} shows the relationship between $T_{\rm eff}$ and [Fe/H] of the FGKM dwarfs ($T_{\rm eff}<7200~\rm K$) hosting close-in giant planets discovered so far\footnote[1]{Data in figures~\ref{fig:Teff-metal}, \ref{fig:Teff-metal_Mdwarfs}, and \ref{fig:metal-Mp} are taken from the NASA Exoplanet Archive  on 2023 Apr. 9 (\url{https://exoplanetarchive.ipac.caltech.edu/}).}.
It shows that M dwarfs with close-in giant planets tend to have high metal abundances, which is qualitatively consistent with the prediction for close-in giant planet formation from the core accretion theory (e.g., \cite{Ida+2004}). 
In addition, it may suggest that the formation of close-in giant planets orbiting low-mass stars (e.g., late K$-$M dwarfs) depend more stringently on stellar metallicity than that around solar-type stars, as also pointed out by \citet{Gan+2022}.
Figure~\ref{fig:Teff-metal} also shows that TOI-519 has the lowest effective  temperature among the stars with close-in giant planets. Although the stellar mass is very low (see table \ref{tab:stellar parameter}), the high metallicity of TOI-519 could be responsible for the formation of the giant planet in the core accretion scheme.
Figure~\ref{fig:Teff-metal_Mdwarfs} shows the relationship between $T_{\rm eff}$ and [Fe/H] of the M dwarfs with close-in transiting planets discovered so far\footnotemark[1]. This figure  shows that most M dwarfs hosting planets of radius greater than about $3~R_\oplus$ are metal-rich, which is also consistent with the recent findings (e.g., \cite{Hirano+2018}; \cite{Chen+2018}). 
It should be noted that stellar metallicities from the NASA Exoplanet Archive are not uniformly derived and different techniques or instruments may present offset/discrepancies (e.g., \cite{Petigura+2018, Passegger+2022}). 

We note that several giant planets with long orbital periods have been discovered even around metal poor M dwarfs (e.g., GJ 832 b, \cite{Bailey+2009}; GJ 3512 b, \cite{Morales+2019}) by the RV method, which may indicate that long-period giant planets are not necessarily formed through the core accretion process.


The distribution of [Fe/H] and $M_p$ is shown in figure~\ref{fig:metal-Mp}\footnotemark[1].
It suggests that close-in giant planets around M dwarfs tend to be less massive compared to those around FGK dwarfs, which is also suggested in \citet{Gan+2022}; in fact, only one planet is heavier than Jupiter. It is theoretically predicted that the final mass of gas giant planets are related to the initial masses of protoplanetary disks (e.g., \cite{Tanigawa+2007}). One possible explanation is that around M dwarfs, there is a lack of gas in the protoplanetary disks that is needed to make a more massive planet. 

The above mentioned correlation between planet occurrence and stellar effective temperature, mass or metallicity, and the trend in planetary mass relative to the host star's mass need to be confirmed through a statistical study by increasing the sample size. Because many candidates of gas giant planets around M dwarfs have been discovered by the TESS survey, more giant planets are expected to be confirmed in the coming years.

\section{Conclusion}

We report the mass determination of TOI-519 b ($M_p = 0.463^{+0.082}_{-0.088}~M_{\rm Jup}$) which is the thirteenth close-in gas giant planet that orbits an M dwarf. The host star, TOI-519, has the lowest effective temperature ($T_{\rm eff} = 3322 \pm 49~\rm K$) among the stars with such planets and has a high metal abundance ([Fe/H]$ = 0.27 \pm 0.09$ dex). The inferred core mass from the simulated thermal evolution of TOI-519 b indicates that it can form in both core accretion and disk instability. 
We report a potential dependence of close-in giant planets formation around low-mass stars on the host star's metallicity, and the trend that they are less massive. 
More samples are needed to confirm the correlation between planet formation and stellar properties, and how it differs from that around the FGK dwarfs.  





\begin{ack}
This research is based on data collected at the Subaru Telescope, which is operated by the National Astronomical Observatory of Japan. We are honored and grateful for the opportunity of observing the Universe from Maunakea, which has the cultural, historical and natural significance in Hawaii.
Data analysis was in part carried out on the Multi-wavelength Data Analysis System operated by the Astronomy Data Center (ADC), National Astronomical Observatory of Japan.
Some of the observations and analyses in this paper were supported by NASA Exoplanet Research Program (XRP) award 80NSSC20K0250. This paper includes data acquired with the Infrared Telescope Facility, which is operated by the University of Hawaii under contract 80HQTR19D0030 with the National Aeronautics and Space Administration.
This work is partly supported by JST CREST Grant No. JPMJCR1761, JSPS KAKENHI Grant No. JP15H02063J, JP17H04574, JP18H05439, JP18H05442, JP19K14783, JP20K14518, JP20K14521, JP21H00035, JP21K13955, JP21K13975, JP21K20376, and JP21K20388, Grant-in-Aid for JSPS Fellows Grant No. JP20J21872, MEXT KAKENHI Grant No. 22000005, and Astrobiology Center.

\end{ack}





\bibliography{ref}{}
\bibliographystyle{aasjournal}

\end{document}